\documentclass[aps,prl,twocolumn,superscriptaddress,longbibliography]{revtex4-2}

\usepackage{amsmath}
\usepackage{amssymb}
\usepackage{mathtools}
\usepackage{graphicx}
\usepackage{xcolor}
\usepackage{hyperref}
\usepackage{tabularx}
\usepackage{svg}
\usepackage{subfigure}

\usepackage{tikz}
\usetikzlibrary{arrows.meta, 3d, calc, positioning}

\newcommand{\ket}[1]{|#1\rangle}
\newcommand{\bra}[1]{\langle #1|}

\newcommand{\Tr}{\mathrm{Tr}}

\newcommand{\Mb}{\mathcal{M}_b}

\hypersetup{
    colorlinks=true,
    linkcolor=blue,
    citecolor=blue,
    urlcolor=blue
}

\begin{document}

\title{The Bott Metric: A Real-Space Bridge Between Topology and Quantum Metric} 

\author{Kaustav Chatterjee}
\email{kaustavc@iisc.ac.in}
\affiliation{Solid State and Structural Chemistry Unit, Indian Institute of Science, Bangalore 560012, India}

\author{Ronika Sarkar}
\affiliation{Solid State and Structural Chemistry Unit, Indian Institute of Science, Bangalore 560012, India}
\affiliation{Department of Physics, Indian Institute of Science, Bangalore 560012, India}

\author{Md Afsar Reja}
\affiliation{Solid State and Structural Chemistry Unit, Indian Institute of Science, Bangalore 560012, India}

\author{Awadhesh Narayan}
\email{awadhesh@iisc.ac.in}
\affiliation{Solid State and Structural Chemistry Unit, Indian Institute of Science, Bangalore 560012, India}

\begin{abstract}
The Bott index has become an indispensable tool to probe the topology of quantum matter, particularly in systems lacking translational symmetry. Constructed from a plaquette operator, it retains the phase information while discarding the amplitude. Here we introduce and develop the \emph{Bott metric}, which captures this complementary amplitude information and provides a measure of the underlying quantum metric of the system. We show that, in the thermodynamic limit, the Bott metric converges to the trace of the integrated quantum metric. Our framework provides a new route to reveal the quantum metric structure in non-periodic systems, which we illustrate using representative examples ranging from disordered to amorphous models. More broadly, our definition of the Bott metric unifies the notion of topological invariants and quantum metric under the same overarching plaquette operator construction.
\end{abstract}

\maketitle

\section{Introduction}
\label{sec:intro}

Topological invariants such as the Chern number, encoded by the Berry curvature, are a cornerstone of modern band theory and topological phases of matter~\cite{berry1984quantal,Thouless:1982}. However, this curvature is only the imaginary part of a broader object, the quantum geometric tensor, whose real part defines the quantum metric~\cite{ProvostVallee1980,ReviewQGTSept,verma2026quantum}. The quantum metric quantifies the distance between quantum states, and its integral defines the integrated quantum metric (IQM). Beyond being a geometric descriptor, the IQM enters directly into measurable response -- it controls the quantum metric contributions to superconducting stiffness in flat and topological bands, and underpins associated bounds~\cite{FirstPavi,AnotherPavi,boundedMATBGBernevig}. In the modern theory of insulators, the same integrated metric content appears in real-space localization diagnostics. It is tied to the gauge-invariant part of Wannier localization~\cite{MarzariVanderbilt1997,localtheoryoftheinsulating}, and can be viewed through ground-state position-space fluctuations~\cite{Resta2011}. These fluctuations enter optical sum rules via the fluctuation-dissipation relations~\cite{SWMsumrule}. Recent work has leveraged this connection between topology, IQM, and optical absorption to derive fundamental bounds on excitation gaps in Chern insulators~\cite{onishi2024fundamental}.

Despite these advances, accessing quantum geometry in systems without translational symmetry remains challenging. In real materials and engineered platforms alike, disorder and aperiodicity are ubiquitous, yet topological phases and their quantized responses can remain robust. Capturing such phases without momentum-space structure requires real-space formulations. In this context, Loring and Hastings introduced the Bott index, an integer-valued real-space invariant, that can be evaluated directly in finite systems~\cite{Hastings_Loring:2010,LoringHastingsEPL2010}. Their construction combines the occupied-state projector with boundary-condition twists along the two spatial directions to define a plaquette operator. In the thermodynamic limit, the phase accumulated by this operator converges to $ 2\pi $ times the Chern number~\cite{Hastings_Loring:2010,Prodan_review:2011,LoringHastingsEPL2010}. As a result, the Bott index provides a powerful and practical diagnostic of topology, and has been widely applied to identify topological phases in disordered, quasicrystalline, amorphous and time-dependent systems~\cite{Huang_2018,GrushinTopomarkeramorphous,Toniolo:2018}.

Here we introduce the concept of the Bott metric, a real-space probe of the quantum metric obtained from the magnitude of the plaquette operator, complementing the Bott index, which captures its global phase. The Bott metric provides direct access to quantum distances in finite systems without invoking translational symmetry, thereby overcoming a central limitation of existing approaches. We demonstrate that, in the thermodynamic limit, it converges to the trace of the IQM, establishing a direct correspondence between real-space constructions and momentum-space quantum metric. Our results place quantum metric and topological information on equal footing within a single plaquette-operator framework, revealing a unified spectral structure that encodes quantum metric and topology.

\begin{figure*}[t]
    \centering
    \includegraphics[width=\textwidth]{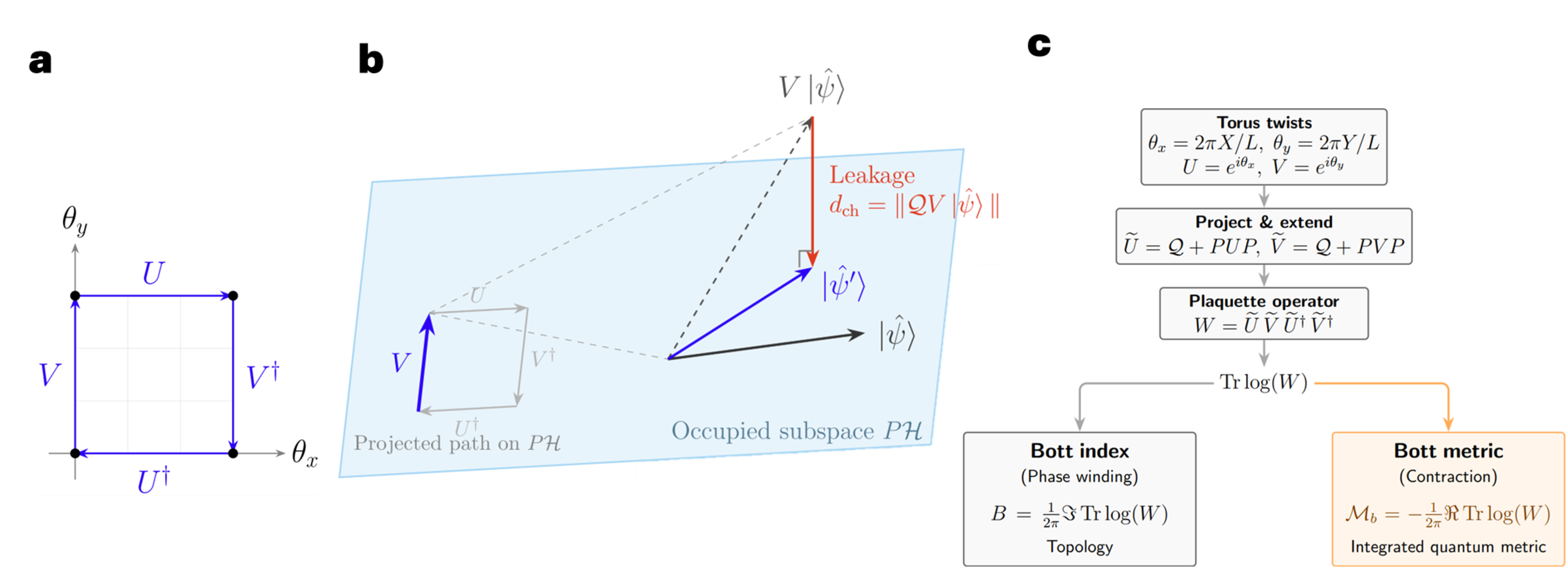}
    \caption{
    \textbf{Plaquette operator framework leading to the Bott metric.}
    (\textbf{a}) Plaquette formed by the torus twists $U$, $V$, $U^\dagger$, and $V^\dagger$ in twist-angle space. While the unprojected loop closes trivially, its compression to the occupied subspace produces a nontrivial response.
    (\textbf{b}) A single projected step on the occupied subspace $P\mathcal H$. Acting with a twist operator (shown here as $V$) generally pushes a normalized occupied state $\ket{\hat\psi}$ slightly outside $P\mathcal H$. Projecting back removes the component in $\mathcal Q\mathcal H$, whose norm $d_{\mathrm{ch}}=\|\mathcal Q V\ket{\hat\psi}\|$ is the one-step leakage. In the small-twist limit, this leakage gives the elementary contribution to the quantum distance.
    (\textbf{c}) Summary of the full construction. From the torus twists $U=e^{i\theta_x}$ and $V=e^{i\theta_y}$, with $\theta_x=2\pi X/L$ and $\theta_y=2\pi Y/L$, we form the projected-and-extended operators $\widetilde U=\mathcal Q+PUP$ and $\widetilde V=\mathcal Q+PVP$, and the plaquette operator $W=\widetilde U\widetilde V\widetilde U^\dagger\widetilde V^\dagger$. The quantity $\Tr\log(W)$ then yields two complementary diagnostics: its imaginary part gives the Bott index, while its real part, introduced below as the Bott metric, measures the total loop contraction and, in the localized regime, yields the integrated quantum metric.
    }
    \label{fig:visualandtable}
\end{figure*}

\section{Formalism}
\label{sec:formalism}

We consider a single-particle Hamiltonian $H$ acting on the total single-particle Hilbert space $\mathcal{H}$ of a finite two-dimensional system with periodic boundary conditions (a torus) of linear size $L$ and area $A=L^2$ . A convenient concrete realization is $\mathcal{H}=\mathrm{span}\{|\mathbf{r},\alpha\rangle\}$, where $\mathbf{r}$ labels lattice sites (or discretized positions on the torus) and $\alpha$ collects internal degrees of freedom (orbitals/spin/sublattice). We place the Fermi energy, $E_F$, in a spectral gap. In disordered or aperiodic settings, we instead assume a mobility-gap regime in which the associated Fermi projector remains local in the real space~\cite{Prodan_review:2011,Bellissard:1994,ReviewLocalization}. We denote the projector onto occupied states by

\begin{equation}
P := \chi_{(-\infty,E_F]}(H)=\sum_{\varepsilon_n\le E_F}\ket{n}\bra{n},\qquad \mathcal{Q}:=\mathbb{I}-P,
\label{eq:projector}
\end{equation}

where $\chi_{(-\infty,E_F]}$ is the indicator (step) function. Here, $\varepsilon_n$ and $\ket{n}$ are the eigenvalues and eigenvectors of $H$, respectively, and $\mathbb{I}$ denotes the identity on $\mathcal{H}$. The projector $\mathcal{Q}$ projects onto the complementary (unoccupied) subspace. We write $P\mathcal{H}$ and $\mathcal{Q}\mathcal{H}$ for the occupied and unoccupied subspaces of $\mathcal{H}$, respectively.

\subsection{Projected twists, leakage, and geometric distance} \label{subsec:proj_translations}

To probe the occupied-subspace response to torus twists, we introduce the real-space phase (twist) operators

\begin{equation}
U:=e^{i\theta_x},\qquad V:=e^{i\theta_y},\qquad \theta_{x,y}=\frac{2\pi}{L}r_{x,y},
\end{equation}

where $r_x:=X$ and $r_y:=Y$ are the position operators on the finite torus. We then restrict these twists to the occupied subspace by defining

\begin{equation}
U_P:=PUP,\qquad V_P:=PVP.
\end{equation}

We also extend these projected twists to operators on the full Hilbert space by letting them act as the identity on the unoccupied sector, namely, $\widetilde U := \mathcal{Q} + PUP$ and $\widetilde V := \mathcal{Q} + PVP$~\cite{Hastings_Loring:2010,LoringHastingsEPL2010,toniolo2022bott}.

Concretely, on $P\mathcal H$ we have 

\begin{equation} 
\|U_P\ket{\psi}\|=\|PU\ket{\psi}\|\le \|U\ket{\psi}\|=\|\ket{\psi}\|, 
\end{equation} 

for every $\ket{\psi}\in P\mathcal H$, and similarly for $V_P$. We refer to such operators as \emph{contractions} on $P\mathcal H$, meaning operators $A$ satisfying $\|A\|\le 1$ (equivalently $A^\dagger A\le P$). For a normalized $\ket{\psi}\in P\mathcal H$, we can decompose the twisted state as 

\begin{equation} 
U\ket{\psi}=PU\ket{\psi}+\mathcal{Q}U\ket{\psi}, 
\end{equation} 

and orthogonality gives the exact norm decomposition 

\begin{equation} 
1=\|PU\ket{\psi}\|^2+\|\mathcal{Q}U\ket{\psi}\|^2. 
\label{eq:pythagorean_leakage} 
\end{equation} 

We call $\|\mathcal{Q}U\ket{\psi}\|^2$ the \emph{leakage} generated by one projected twist step -- it is the norm shed into the unoccupied sector. If $PU\ket{\psi}\neq 0$, we define the normalized post-projection state $\ket{\psi_{\mathrm{proj}}}:=PU\ket{\psi}/\|PU\ket{\psi}\|$ Equation~\eqref{eq:pythagorean_leakage} then gives us

\begin{equation} 
\|\mathcal{Q}U\ket{\psi}\|^2 =1-\bigl|\bra{\psi}U^\dagger\ket{\psi_{\mathrm{proj}}}\bigr|^2. 
\label{eq:FS_chordal_leakage} 
\end{equation} 

Here the right-hand side is the squared chordal Fubini-Study distance between the ray of $U\ket{\psi}$ and the ray of its normalized projection back into $P\mathcal H$~\cite{ProvostVallee1980,BengtssonZyczkowski2017,BrodyHughston2001}. For a twist step of $2\pi/L$ the distance is quadratic in the step size, with the leading term given by the quantum metric~\cite{ProvostVallee1980,WGTmanifold}. Thus, a single projected twist step defines a mismatch in quantum distance between the twisted state and its normalized post-projection state. We next compose four such projected steps into a plaquette loop on the torus and quantify the net contraction accumulated around the loop.

\subsection{Plaquette operator and accumulated contraction}
\label{subsec:bott_loop_metric}

We now combine the twists into a closed loop on the torus. The product $UVU^\dagger V^\dagger$ corresponds to twisting along $x$ and $y$, followed by undoing these twists, and it is therefore natural to interpret it as a plaquette loop. We define the plaquette operator on the full space and the projected space as

\begin{equation}
W:=\widetilde U\,\widetilde V\,\widetilde U^\dagger \widetilde V^\dagger,\qquad 
W_P:=PWP=U_PV_PU_P^\dagger V_P^\dagger.
\label{eq:Wp_def_main}
\end{equation}

Since $\widetilde U$ and $\widetilde V$ act as the identity on $\mathcal Q\mathcal H$, one has $W=\mathbb I_{\mathcal Q}\oplus W_P$, so the nontrivial loop action is entirely encoded by $W_P$ on $P\mathcal H$.

To see how the loop accumulates leakage, it is convenient to follow one state along the plaquette. Let $\ket{\psi_0}\in P\mathcal H$ and we then write the loop as four projected steps $\ket{\psi_k}=P U_k\ket{\psi_{k-1}}$ with $U_k\in\{V^\dagger,U^\dagger,V,U\}$. We then define the normalized intermediate states as $\ket{\hat\psi_{k-1}}=\ket{\psi_{k-1}}/\|\psi_{k-1}\|$ and the leakage amplitudes as $d_{\mathrm{ch},k}:=\|\mathcal{Q}U_k\ket{\hat\psi_{k-1}}\|$. From Eq.~\eqref{eq:pythagorean_leakage}, each step reduces the norm by

\begin{equation}
\|\psi_k\|^2=\|\psi_{k-1}\|^2\,(1-d_{\mathrm{ch},k}^2).
\end{equation} 

A schematic of a single step is shown in Fig.~\ref{fig:visualandtable}(b). After loop traversal the net norm-retention factor is

\begin{equation}
\frac{\|W_P\ket{\psi_0}\|^2}{\|\psi_0\|^2}=\prod_{k=1}^4(1-d_{\mathrm{ch},k}^2).
\end{equation}

A particularly transparent case occurs when $\ket{\psi_0}$ is an eigenvector of $W_P$ with eigenvalue $\lambda$, so that $W_P\ket{\psi_0}=\lambda\ket{\psi_0}$ and the left-hand side equals $|\lambda|^2$. In that case, we have

\begin{equation}
|\lambda|^2=\prod_{k=1}^4(1-d_{\mathrm{ch},k}^2),
\end{equation}

such that $|\lambda|<1$ directly measures how much norm is lost after going once around the plaquette. Taking logarithms turns the product of stepwise retention factors into a sum,

\begin{equation}
2\log|\lambda|
=\sum_{k=1}^{4}\log\!\left(1-d_{\mathrm{ch},k}^{2}\right)\le 0,
\label{eq:lambda_logsum_14}
\end{equation}

and therefore at sufficiently large $L$

\begin{equation}
-\log|\lambda|
=-\frac{1}{2}\sum_{k=1}^{4}\log\!\left(1-d_{\mathrm{ch},k}^{2}\right)
\approx \frac{1}{2}\sum_{k=1}^{4} d_{\mathrm{ch},k}^{2}.
\label{eq:lambda_contraction_15}
\end{equation}

In the large-$L$ regime the twist step is $\kappa=2\pi/L$, and in the gapped (or localized) setting each projected step loses only a small amount of norm ($d_{\mathrm{ch},k}^2\ll 1$), which allows the approximation $\log(1-x)\approx -x$.

We can now summarize the overall contraction of the occupied space by a single scalar that multiplies the loop’s contraction factors across all independent directions in $P\mathcal H$. Concretely, for a linear map on the finite-dimensional space $P\mathcal H$, the determinant measures how the map rescales an ${m}$-dimensional volume element (the hypervolume spanned by $m=\mathrm{rank}(P)$ independent vectors). Thus $|\det(W_P)|$ is the basis-independent net volume-retention factor after one loop, and $-\log|\det(W_P)|$ is the corresponding log-volume contraction.  Writing $\{\lambda_j\}$ for the eigenvalues of $W_P$ (counted with algebraic multiplicity), we can express it as

\begin{equation}
-\log|\det(W_P)|
=-\sum_j \log|\lambda_j|
=-\Re\,\Tr\log(W_P).
\label{eq:sum_eigs_trace_16}
\end{equation}

This trace-level quantity is the loop contraction we will use next to define the Bott metric.

\subsection{Defining the Bott metric}
\label{subsec:Bmetric_def}

The Bott index is obtained from the same loop by keeping only the phase information, i.e., the imaginary part of the loop trace-log. The amplitude information of this quantity remains unexplored so far. We retain the amplitude content encoded in the real part and use it to define the Bott metric, $\Mb$, as

\begin{equation}
\Mb:=-\frac{1}{2\pi}\,\Re\,\Tr\log(W).
\label{eq:Bmetric_def_main}
\end{equation}

This definition is unambiguous whenever $W$ is invertible. In this case the real part depends only on the modulus of the determinant and is therefore insensitive to the branch choice of the matrix logarithm,

\begin{equation*}
\Re\,\Tr\log(W)=\log|\det(W)|.
\end{equation*}

To connect Eq.~\eqref{eq:Bmetric_def_main} to the occupied-sector loop, note that $\widetilde U$ and $\widetilde V$ act trivially on $\mathcal Q\mathcal H$ and coincide with $U_P$ and $V_P$ on $P\mathcal H$. Thus, they are block diagonal with respect to $P\oplus\mathcal Q$, which implies $W=\mathbb I_{\mathcal Q}\oplus W_P$ and hence $\det(W)=\det(W_P)$.

Therefore, whenever $W$ is invertible,

\begin{equation*}
\Re\,\Tr\log(W)=\log|\det(W_P)|=\Re\,\Tr\log(W_P).
\end{equation*}

Equivalently, writing $\{\lambda_j\}$ for the eigenvalues of $W_P$ (counted with algebraic multiplicity),

\begin{equation}
\Mb=-\frac{1}{2\pi}\log|\det(W_P)|
=-\frac{1}{2\pi}\sum_j \log|\lambda_j|.
\label{eq:Mb_det_eigs}
\end{equation}

In other words, up to the prefactor $1/(2\pi)$, the Bott metric is the log-volume contraction of the plaquette loop introduced in Eq.~\eqref{eq:sum_eigs_trace_16}.

\subsection{Connecting contraction to integrated quantum metric}
\label{subsec:why_TrG}

To connect loop contraction to quantum metric, we work in the regime where the Fermi projector remains local in real space, which holds both in spectrally gapped systems and in mobility-gap regimes~\cite{Prodan_review:2011,Bellissard:1994,ReviewLocalization}. In this regime, the contraction encoded in $\Re\,\Tr\log(W)$ is controlled by the IQM tensor \cite{romeral2025scaling,Resta2011}

\begin{equation}
G_{\alpha\beta}:=-\frac{2\pi}{A}\,\Re\,\Tr\!\left(P[r_\alpha,P][r_\beta,P]\right),
\label{eq:G_def_main}
\end{equation}

where $\alpha,\beta \in \{x,y\}$. The key simplification is specific to the adjoint plaquette ordering $UVU^\dagger V^\dagger$: the modulus of the loop determinant factorizes exactly, so the real trace-log splits into independent $x$- and $y$-contributions,

\begin{equation}
\Re\,\Tr\log(W_P)
=\Tr\log(U_P^\dagger U_P)+\Tr\log(V_P^\dagger V_P).
\label{eq:additive_contraction}
\end{equation}

A controlled small-twist expansion then shows that, under the locality and thermodynamic-limit assumptions stated in the supplementary information~\cite{suppmaterial}, the Bott metric reproduces the integrated quantum metric in the thermodynamic limit,

\begin{equation}
\lim_{L\to\infty}\Mb=\Tr(G).
\label{eq:Bmetric_to_TrG_main}
\end{equation}

In the supplementary information we prove this equality through a finite-volume periodic formulation and provide the corresponding error control~\cite{suppmaterial}; in particular, in the mobility-gap setting the thermodynamic identification uses both projector locality and the additional thermodynamic-compatibility assumption stated there. Near topological or localization transitions, however, the locality underlying this expansion can fail. In that regime $\Mb$ can become strongly enhanced, reflecting the breakdown of perturbatively norm-preserving projected transport. We further provide a mechanism-based finite-volume singularity criterion for this blow-up~\cite{suppmaterial}. This behavior is consistent with recent real-space studies of the IQM, which also find strong growth near delocalization and thermodynamic divergence in metallic regimes~\cite{romeral2025scaling}. The conceptual relation between the established Bott index and our proposed Bott metric is summarized in Fig.~\ref{fig:visualandtable}(c).

\section{Applications}
\label{sec:numerics}

After introducing the Bott metric, we next illustrate its applications in clean and disordered Qi-Wu-Zhang (QWZ) model and then use it as an independent geometric probe in an amorphous Chern insulator model.

\begin{figure}
   \centering
   \includegraphics[width=\linewidth]{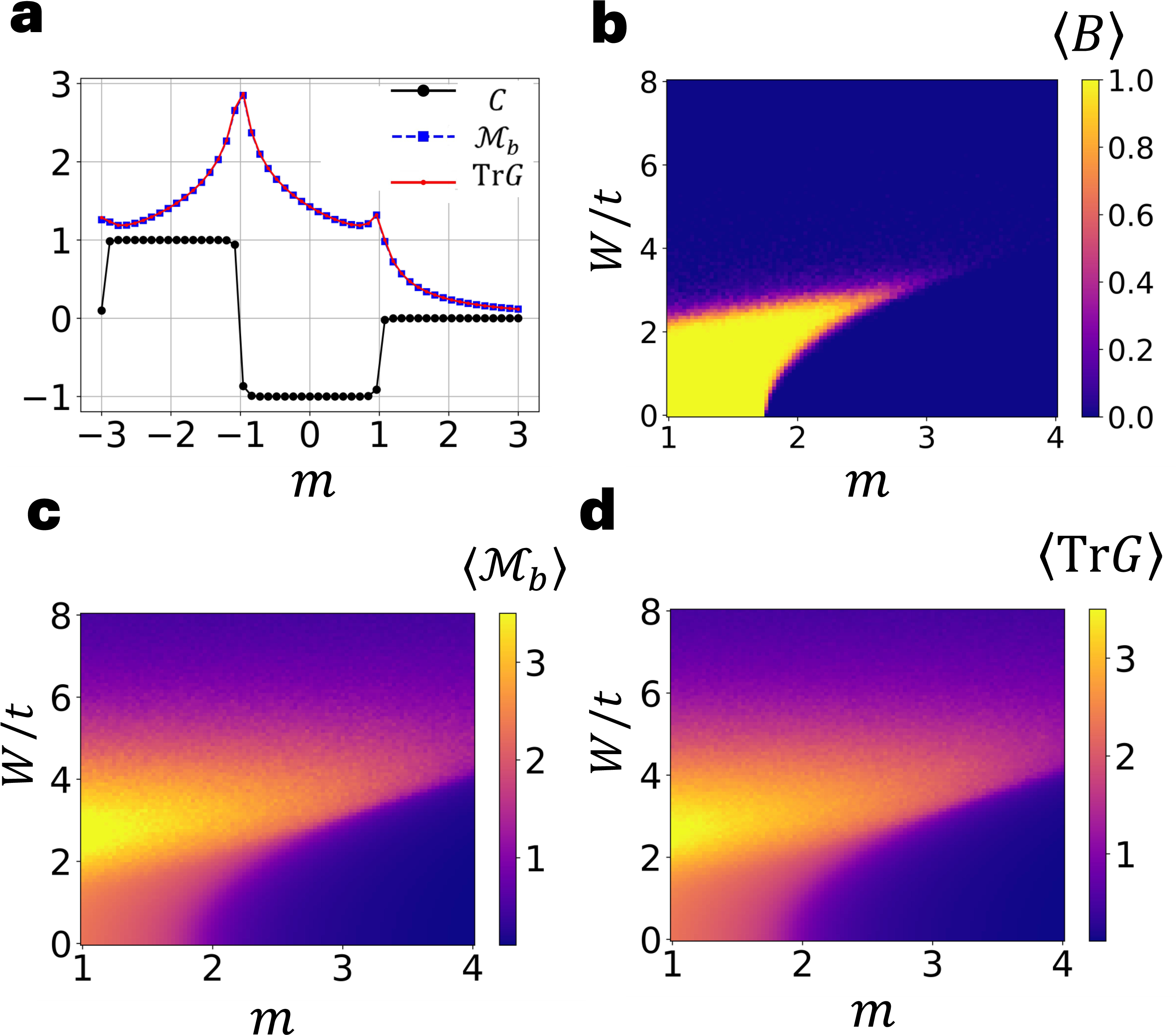}
   \caption{\textbf{Bott metric in clean and disordered Qi-Wu-Zhang model.} \textbf{(a)} The Chern number $C$, the Bott metric $\Mb$, and the integrated quantum metric $\Tr(G)$ as a function of the mass $m$ for the clean Qi-Wu-Zhang model. Here $L=30$ torus was used at half filling ($E_F=0$), with $A=1$ and $B=0.5$. \textbf{(b)} The disorder-averaged Bott index $\langle B\rangle$ over the $(m,W)$ plane on an $L=16$ torus with $t=\nu=1$ and $\gamma=0$ for the disordered Qi-Wu-Zhang model with disorder strength $W$. \textbf{(c)} Disorder-averaged Bott metric $\langle \Mb\rangle$ and \textbf{(d)} integrated quantum metric $\langle \Tr(G)\rangle$ for the same $(m,W)$ scan. In both the clean and the disordered phase diagrams, $\Mb$ closely tracks $\Tr(G)$: the curves in (a) are nearly indistinguishable and, in (c) and (d), both the moderate valued magenta colored region inside the quantized-$\langle B\rangle$ plateau and the bright yellow boundary band occur at the same $(m,W)$ locations and have identical color gradients, demonstrating the equivalence between the Bott metric and the trace of integrated quantum metric.}
\label{fig:qwz}
\end{figure}

\subsection{Clean and Disordered QWZ model}
\label{subsec:qwz_benchmarks}

We first compare the Bott metric $\Mb$ against the conventional integrated quantum metric trace $\Tr(G)$ in the clean case. We consider the celebrated QWZ model on an $L\times L$ square lattice under periodic boundary conditions with two orbitals per site at half filling (see Methods for details). The Hamiltonian reads~\cite{QiWuZhang2006}

\begin{equation}
H=\sum_{\mathbf r} c^\dagger_{\mathbf r}(m+2B)\sigma_z c_{\mathbf r}
+\sum_{\mathbf r}\Big(c^\dagger_{\mathbf r+\hat x}T_xc_{\mathbf r}+c^\dagger_{\mathbf r+\hat y}T_yc_{\mathbf r}+{\rm H.c.}\Big),
\end{equation}

with $T_x=-B\sigma_z+\tfrac{iA}{2}\sigma_x$ and $T_y=-B\sigma_z+\tfrac{iA}{2}\sigma_y$. In the clean QWZ model, $\Mb$ is controlled by how strongly the plaquette steps mix the occupied and unoccupied sectors. Deep in a gapped phase, the occupied subspace $P\mathcal H$ is well separated from $\mathcal Q\mathcal H$, so $\Mb$ remains smooth and finite. As the bulk gap closes at the topological transition points $m=\pm1$, this separation collapses and the mixing becomes strong, leading to sharp peaks in $\Mb$. Accordingly, Fig.~\ref{fig:qwz}(a) shows that $\Mb$ and $\Tr(G)$ closely track each other and develop pronounced cusps at $m=\pm1$, coincident with the jumps in the noncommutative Chern number $C$. We focus on these two transitions; an additional closing at $m=-3$ lies at the edge of the plotted window. The gap-closing states that alter the topology also generate a singular metric response, signaling an incipient delocalized character, arising from metallic behavior at the gap closing points. The near perfect agreement between $\Mb$ and $\Tr(G)$ shows that the Bott metric captures this critical geometric response directly in real space.

We next consider the disordered QWZ model (see Methods for details of the disordered model). Across the mobility-gap regime, at half filling, $\langle B\rangle$ remains quantized [Fig.~\ref{fig:qwz}(b)], while $\langle \Mb\rangle$ measures how strongly the phase-step-and-project procedure mixes $P\mathcal H$ with $\mathcal Q\mathcal H$ across the sample (through the trace). Figs.~\ref{fig:qwz}(c) and (d) show the disorder-averaged $\langle \Mb\rangle$ and $\langle \Tr(G)\rangle$ in the same $(m,W/t)$ sweep as the Bott-index phase diagram in Fig.~\ref{fig:qwz}(b). In the low-disorder topological region where $\langle B\rangle\approx 1$ (roughly $m\sim 1$--$2$ and $W/t\sim 0$--$2$), both $\langle \Mb\rangle$ and $\langle \Tr(G)\rangle$ are moderate (greenish), consistent with localized states near $E_F$ and hence weak $P\mathcal H$--$\mathcal Q\mathcal H$ mixing under the plaquette steps. 

Approaching the phase boundary where the $\langle B\rangle\approx 1$ plateau terminates, both quantities develop a bright ridge (yellow) at intermediate disorder, concentrated around $W/t\sim 2$--$4$, indicating reduced localization near $E_F$ and a sharp increase in the effectiveness of $P\mathcal H$--$\mathcal Q\mathcal H$ mixing. As $m$ increases, this ridge weakens and shifts in tandem with the shrinking topological region in Fig.~\ref{fig:qwz}(b). The ridge location and overall pattern closely coincide between Figs.~\ref{fig:qwz}(c) and (d), showing that $\Mb$ and $\Tr(G)$ remain in good agreement under disorder.

\begin{figure}[t]
    \centering
    \includegraphics[width=\linewidth]{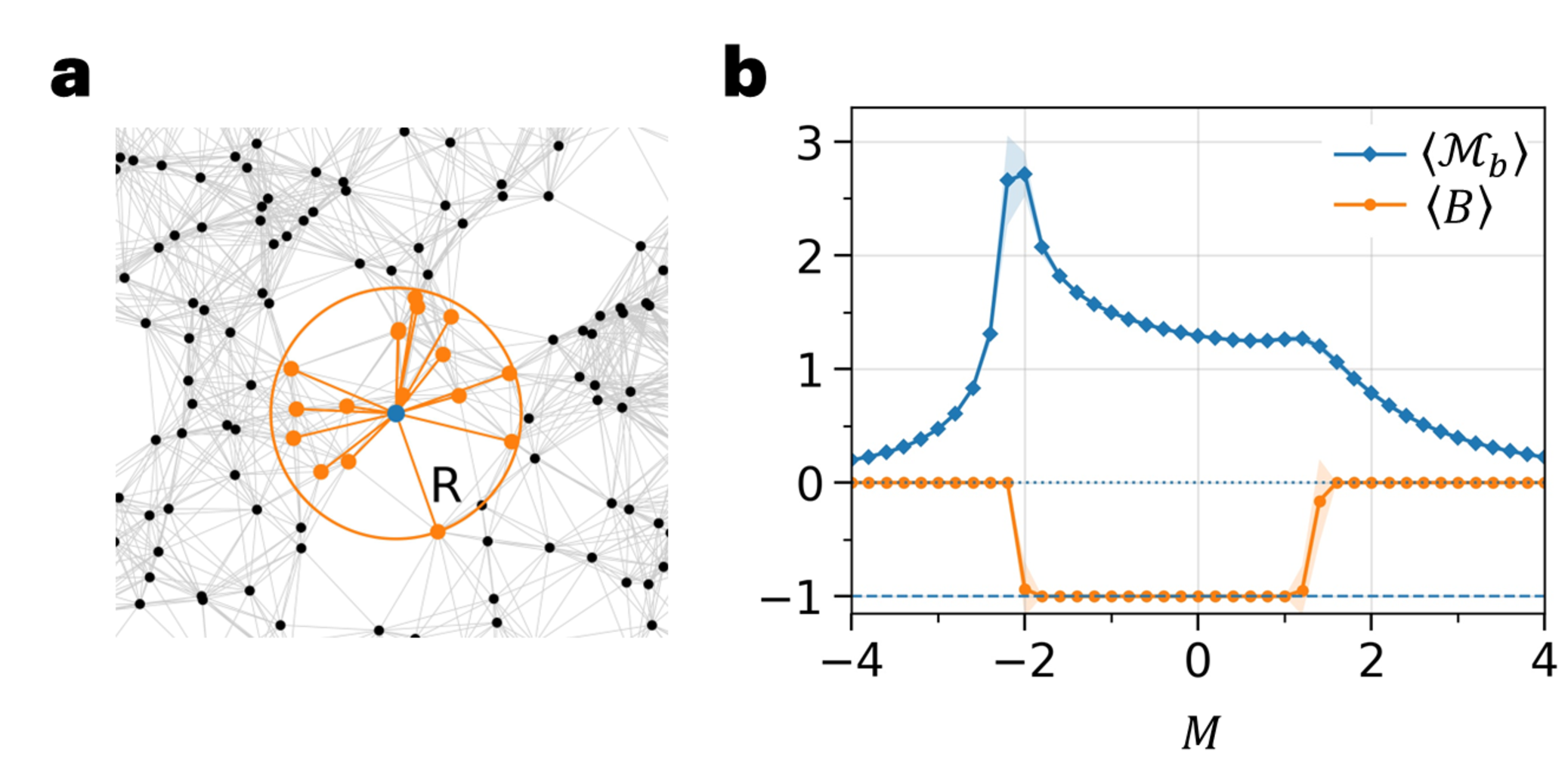}
    \caption{\textbf{Bott metric in amorphous systems.} \textbf{(a)} Representative realization of an amorphous model with a hopping cutoff radius $R$. \textbf{(b)} Ensemble-averaged Bott index $\langle B\rangle$ and the Bott metric $\langle \Mb\rangle$ versus the mass parameter $M$. The shaded portions indicate the standard error of the mean obtained from the ensemble averaging. While $\langle B\rangle$ identifies a broad topological plateau, $\langle \Mb\rangle$ varies strongly across the same window and exhibits an asymmetric peak structure at the two topological phase transition points, revealing additional localization-sensitive geometric information beyond the quantized topological label. Here we set $L_x=L_y=25$, $\rho=0.6$, $R=4.0$, $t_2=0.25$, and $\lambda=0.5$.}
    \label{fig:amorphous}
\end{figure}

\subsection{Amorphous Chern insulator}
\label{subsec:amorphous_app}

The characterization of topological amorphous solids is intrinsically challenging. In the absence of translational symmetry the topology and quantum metric of the system must be investigated using genuinely real-space probes~\cite{Mitchell2018,Corbae2023,MunozSegovia2023}. To showcase our Bott metric approach, we next turn to the amorphous Chern insulator model of Agarwala and Shenoy~\cite{AgarwalaShenoyPRL2017}, as schematically shown in Fig.~\ref{fig:amorphous}(a) (see Methods for details). 

In Fig.~\ref{fig:amorphous}(b), we show the average Bott index $\langle B\rangle$, which is quantized over a broad window of $M$, while $\langle \Mb\rangle$ varies substantially across the same topological plateau. This variation has a direct localization interpretation -- in a mobility-gap regime, localization is reflected in the locality of the occupied projector $P$, and this locality suppresses the transfer of weight from $P\mathcal H$ into $\mathcal Q\mathcal H$ under the plaquette steps. Accordingly, larger values of $\langle \Mb\rangle$ signal an enhanced $P\mathcal H$--$\mathcal Q\mathcal H$ mixing and a stronger delocalization tendency near $E_F$. On the other hand, the decline of $\langle \Mb\rangle$ deeper inside the plateau is consistent with progressively weaker mixing as $P$ becomes more local.

Notably the two plateau edges are visibly inequivalent. Near the negative-$M$ transition ($M\simeq -2$) $\langle \Mb\rangle$ exhibits a sharper, higher peak, while near the positive-$M$ transition ($M\simeq 1$) the feature is much weaker. 
This asymmetry aligns with a finite-size analysis, which finds distinct approaches to the thermodynamic limit at the two gap closings -- the negative-$M$ transition shows a faster closing of the minimum gap ($\sim 1/L^2$), whereas the positive-$M$ transition shows a slower overall trend (gap scaling as $\sim 1/L$) together with pronounced size-dependent oscillations, attributed to interference/rare-region effects~\cite{AgarwalaShenoyPRL2017}. Thus, while $\langle B\rangle$ identifies the topological plateau, $\Mb$ reveals a \emph{hidden} real-space diagnostic of how strongly the occupied sector is destabilized near each critical point, resolving the localization-sensitive asymmetry between the two transitions within the same model.

\section{Discussion and Conclusions}
\label{sec:conclusion}

In this work we have extended the real-space Bott index framework beyond topology by extracting, from the same plaquette operator, a diagnostic of the IQM. The central observation is that projected transport, implemented through successive twist-and-project operations, is intrinsically contractive on the occupied subspace. The Bott metric, $\Mb$, quantifies the resulting cumulative norm leakage around the plaquette, and we show rigorously that, in the thermodynamic limit, it converges to the trace of the IQM, thereby providing an alternative real-space formulation of integrated quantum metric.

A key practical advantage of this construction is its negligible computational and implementation overhead: once a Bott index pipeline is in place, $\Mb$ can be evaluated at essentially no additional cost. This makes the Bott metric an immediately accessible real-space route to the IQM in disordered and aperiodic systems, and readily applicable across the wide range of platforms where the Bott index is already employed -- including topological insulators and superconductors, non-Hermitian systems, amorphous materials, hyperbolic systems, and beyond~\cite{titum2015disorder,FracAmorphous,Huang_2018,AubreNonHerm,Amorphspinbott,Brataas_2020,yu2020topological,kollar2019hyperbolic}. Importantly, the Bott metric remains well defined even in topologically trivial gapped systems, where it serves as a purely quantum metric probe.

Stepping back, the central conceptual advance of this work is a unification: topology and quantum metric emerge as complementary facets of a single spectral object -- the plaquette operator -- with the Bott index capturing its phase and the Bott metric its amplitude. This perspective extends the scope of real-space approaches beyond topological classification, and suggests broader applicability. In particular, it points to natural connections with Wilson loop constructions in other settings, and opens a route towards a broader real-space framework for the metric sector of quantum geometry in systems where momentum-space methods are unavailable.

\section{Methods}
\label{sec:methods}

\subsection*{Clean Qi-Wu-Zhang model}

For benchmarking the Bott metric in clean systems, we use the translationally invariant QWZ Chern insulator model~\cite{QiWuZhang2006} on an $L\times L$ square lattice with periodic boundary conditions and two orbitals per site at half filling. Writing $c_{\mathbf r}=(c_{\mathbf r,\uparrow},c_{\mathbf r,\downarrow})^{\mathsf T}$ and letting $\sigma_{x,y,z}$ act in the orbital space, the Hamiltonian is

\begin{equation}
H=\sum_{\mathbf r} c^\dagger_{\mathbf r}(m+2B)\sigma_z c_{\mathbf r}
+\sum_{\mathbf r}\Big(c^\dagger_{\mathbf r+\hat x}T_xc_{\mathbf r}+c^\dagger_{\mathbf r+\hat y}T_yc_{\mathbf r}+{\rm H.c.}\Big),
\label{eq:H_clean_QWZ_methods}
\end{equation}

with nearest-neighbor hopping matrices

\begin{equation}
T_x=-B\sigma_z+\tfrac{iA}{2}\sigma_x,
\qquad
T_y=-B\sigma_z+\tfrac{iA}{2}\sigma_y.
\label{eq:TxTy_clean_QWZ_methods}
\end{equation}

Here $m$ is the tunable mass parameter and $A,B$ control the inter-orbital mixing and orbital-dependent hopping, respectively. 

\subsection*{Disordered Qi-Wu-Zhang model}

For the disordered benchmark we use the Hermitian limit of the model introduced in Ref.~\cite{TangPRA2020}, following Refs.~\cite{XueProdan2013,Priest2014,Dahlhaus2015,Ulcakar2020},

\begin{equation}
H
=
\sum_{x}\sum_{j=x,y}
\Big(
c_x^\dagger T_j c_{x+\hat e_j}
+
c_{x+\hat e_j}^\dagger T_j^\dagger c_x
\Big)
+
\sum_x c_x^\dagger M_x c_x,
\label{eq:H_disordered_PRA_Hermitian}
\end{equation}

where the nearest-neighbor hopping matrices are

\begin{equation}
T_j
=
-\frac{t_j}{2}\sigma_z
-\frac{i v_j}{2}\sigma_j,
\qquad j\in\{x,y\},
\label{eq:Tj_def}
\end{equation}

and the onsite term is a Zeeman-like mass $M_x = m_x\,\sigma_z$. Here $t_j$ controls the orbital-conserving hopping along direction $j$ and $v_j$ controls the inter-orbital mixing along direction $j$ through $\sigma_j$. In our calculations we take the isotropic choice, without loss of generality, $t_x=t_y=t$ and $v_x=v_y=v$, and set the energy unit by $t=v=1$. Disorder enters through a site-dependent mass

\begin{equation}
m_x = m + W\,\omega_x,
\qquad
\omega_x \sim \mathrm{Unif}[-1,1],
\label{eq:mass_disorder}
\end{equation}

where $m$ is the mean mass and $W$ is the disorder strength. The values of $\omega_x$ are independent across sites and chosen from a uniform distribution.

\subsection*{Amorphous Chern insulator model}

We consider an amorphous Chern insulator model, where $N$ sites are placed in a square box of area $V=L_x L_y$ by an uncorrelated uniform (Poisson) point process with density $ \rho = N/V$~\cite{AgarwalaShenoyPRL2017}. Each site hosts two orbitals. Hopping is restricted to pairs with separation $|\mathbf r_{IJ}| \le R$, where $R$ is a fixed cutoff radius, and distances/angles are computed on the torus defined by the periodic boundary conditions. The Hamiltonian takes the form

\begin{equation}
H = \sum_{I\alpha}\sum_{J\beta} t_{\alpha\beta}(\mathbf r_{IJ})\, c^\dagger_{I,\alpha} c_{J,\beta},
\qquad
t_{\alpha\beta}(\mathbf r) = t(r)\,T_{\alpha\beta}(\hat{\mathbf r}),
\end{equation}

with an onsite term $t_{\alpha\beta}(\mathbf 0)=\epsilon_{\alpha\beta}$. The radial envelope is

\begin{equation}
t(r) = C\,\Theta(R-r)\,e^{-r/a},
\end{equation}

and we use the units $a=1$ with $C=e$, such that $t(1)=1$. The angular dependence is encoded by the matrix $T_{\alpha\beta}(\hat{\mathbf r})$, where $\hat{\mathbf r}$ is specified by the polar angle $\theta$ of $\mathbf r$ in two dimensions. The model parameters are $\rho$ (site density), $R$ (hopping range), and internal couplings $(\lambda,t_2,M)$, where $\lambda$ controls inter-orbital mixing, $t_2$ controls intra-orbital hopping, and $M$ is the mass term that tunes the phase. We work at one fermion per site, i.e., half filling of the two-orbital model.

\subsection*{Open-bulk non-commutative Chern number }

To label the topology in the clean benchmark we compute a real-space Chern number from $P$ using the standard projector-commutator expression. Let $X$ and $Y$ be the coordinate operators, diagonal in the site basis. We evaluate~\cite{Bellissard:1994,ProdanSchulzBaldes}

\begin{equation}
C
=
\frac{2\pi i}{L_x' L_y'}\,
\Tr'\!\Big(
P\big[[X,P],[Y,P]\big]
\Big),
\label{eq:open_bulk_chern}
\end{equation}

where $\Tr'$ denotes the trace restricted to a central window of size $L_x'\times L_y'$ (with a fixed buffer cut away from each edge) to suppress finite-size/boundary effects. In the clean case we compute $C$ for a single configuration.

\subsection*{Bott index}

On the torus one defines the unitary position-phase operators

\begin{equation}
U=\exp\!\Big(\frac{2\pi i}{L_x}X\Big),
\qquad
V=\exp\!\Big(\frac{2\pi i}{L_y}Y\Big).
\end{equation}

Given $P$ and $Q=I-P$, we form the full-space projected operators

\begin{equation}
\widetilde U = Q + PUP,
\qquad
\widetilde V = Q + PVP,
\end{equation}

and the plaquette loop operator (with adjoint plaquette ordering)

\begin{equation}
W = \widetilde U\,\widetilde V\,\widetilde U^\dagger\,\widetilde V^\dagger.
\end{equation}

The Bott index is extracted from the phase of the principal trace-log~\cite{Loring_2019,LoringHastingsEPL2010},

\begin{equation}
B
=
\frac{1}{2\pi}\Im\,\Tr\log(W),
\end{equation}

where $\log(W)$ denotes the principal matrix logarithm, which is evaluated in practice from the eigenvalues of $W$.

\subsection*{Configuration averaging}

A configuration indicates a complete specification of the randomness used to construct the Hamiltonian. For the disordered lattice model above, one configuration is a full set of onsite random variables $\{\omega_x\}$ (hence one realization of $m_x$). For the amorphous model, one configuration is a full random point set and the induced set of bonds. For any observable $O$ (e.g., $B$, $\Mb$, or $C$ when applicable), we compute $O_s$ on each independent configuration $s=1,\dots,N_{\mathrm{cfg}}$, and report the configuration average $\langle O\rangle$ and the standard error of the mean $\mathrm{SEM}(O)$, as

\begin{equation}
\langle O\rangle
=
\frac{1}{N_{\mathrm{cfg}}}\sum_{s=1}^{N_{\mathrm{cfg}}} O_s, \quad \mathrm{SEM}(O)
=
\sqrt{\frac{\mathrm{Var}(O_s)}{N_{\mathrm{cfg}}}}.
\end{equation}

\section{Code Availability}
The codes that support the findings of this study are available from the corresponding authors upon reasonable request.

\section{Data Availability}
The data that support the findings of this study are available from the corresponding authors upon reasonable request.

\section{Acknowledgments}
K.C. acknowledges the support from University Grants Commission (UGC), Government of India under the Senior Research Fellowship (SRF) scheme for this project. R.S. is supported by the Prime Minister's Research Fellowship (PMRF). M.A.R. acknowledges a graduate fellowship of the Indian Institute of Science. A.N. thanks DST MATRICS grant (MTR/2023/000021) for support.

\section{Author Contributions}
K.C. carried out the analysis and calculations with inputs from R.S. and M.A.R. A.N. and K.C. conceived the research. A.N. supervised the project. All authors contributed to the writing of the manuscript.

\section{Competing  interest}
The authors declare no competing interests.

\bibliographystyle{apsrev4-2}
\bibliography{refs}

\end{document}